\documentstyle[twocolumn]{article}

\def\r{\hangindent=1pc  \noindent}
\def\kms{km s$^{-1}$}
\def\deg{$^\circ$}
\def\Deg{^\circ}

\def\tableline{\hline}

\begin{document}
\title{Bipolar hypershell models of the extended hot interstellar medium
in  spiral galaxies}
\author{Yoshiaki Sofue\inst{1} and Andreas Vogler\inst{2}}
   \author{
1. Institute of Astronomy, University of Tokyo, Mitaka, Tokyo 181-0015, Japan\\
(sofue@ioa.s.u-tokyo.ac.jp) \\
2. CEA/Saclay, DAPNIA, Service d'Astrophys, L'Ormes
des Mersiers, B\^at. 709, \\
F-91191 Gif-sur-Yvette, France
}
\maketitle

\begin{abstract}
We simulated the million degree interstellar medium
and its soft X-ray images in the disk and halo of
spiral galaxies using the bipolar hypershell (BHS) model. In this model
dumbbell- or hourglass-shaped expanding shells of
several kpc radii are produced by a sudden energy release
in the central region. We then applied our model to a mini-sample of
starburst galaxies seen under different inclinations, namely the nearly edge-on
galaxies NGC~253, NGC~3079 and M~82, the highly inclined galaxy
NGC~4258 as well as  the nearly face-on galaxies NGC~1808 and M~83.
For all galaxies, our results reproduce the X-ray characteristics observed
in the 0.1--2.4 keV ROSAT energy band:
The bipolar hypershell morphology, the spectral energy distribution of the
diffuse disk and halo emission as well as absorption gaps in the diffuse
X-ray emission caused by a shadowing of soft X-rays due to cold intervening
gas in the disks of the galaxies. In general,
the required total energy for the starburst
is estimated to be of the order of $10^{55}$ ergs,
corresponding to the overall kinetic energy generated in
$\sim 10^4$ type-II supernova explosions.
The expansion velocity of the shells is estimated to be $\sim 200$ \kms,
which is necessary to heat the gas to $\sim 0.2$ keV (2.3 million K),
and the age to be of the order of $3 \times 10^7$ years. In the
case of the very nearby, nearly edge-on galaxy NGC~253
all characteristics of the BHS model can be studied with high
spatial resolution.

Using the property that the shell morphology is sensitive to the
ambient density distribution, we propose using soft X-ray data to
probe the gas distributions in the disk, halo and intergalactic space in
general. The application of our model to
images at higher spatial and spectral resolution, as provided by {\it Chandra}
and XMM, will help us to further
disentangle the ISM density distributions
 and will lead to a better understanding of the disk halo interface.

\vskip 2mm
Keywords {galaxies: general -- galaxies: individual: NGC~253, NGC~1808,
NGC~3079, NGC~4258, M~82 (NGC~3034), the Milky Way -- galaxies: evolution --
galaxies: ISM -- galaxies: starburst
-- galaxies: spiral -- galaxies: structure -- galaxies: X-rays --
X-rays: general -- X-rays: galaxies}

\end{abstract}

\section{Introduction}

A bipolar hypershell (BHS) model has been proposed to explain the
double-horn features observed in the radio-continuum halo of the
starburst galaxy NGC 253 (Sofue 1984).
Recent soft-X-ray observations of NGC 253 with ROSAT
revealed extended outer-halo emission with double-horn
morphology, and is interpreted as being a hypershell driven by a
central starburst (Vogler and Pietsch 1999a; Pietsch et al.  2000).
Many spiral galaxies exhibit similar
galactic-scale outflows in the form of dumbbell-shaped hypershells,
bipolar cylinders, or galactic-scale jets.
Examples of prototypes of such galaxies are M~82 (NGC~3034),
NGC 1808 and NGC 3079
(see next section for references).
Starburst-origin models for extended halo features in
galaxies were proposed by several authors (Heckman et al. 1993;
Tomisaka and Ikeuchi 1987; Suchkov et al. 1994, 1996).

The Milky Way also exhibits bipolar hypershells extending
above and below the galactic plane in soft X-rays
(Snowden et al. 1997) and radio continuum (e.g. Haslam et al. 1982),
which are also interpreted as an effect of a giant $\Omega$-shaped
shock front originating from a starburst some $10^7$ years
ago in the Galactic Center (Sofue 1977, 1984, 1994, 2000).
These hypershells and flows, including
that in the Milky Way, require a similar amount of total energy
of the order of $\sim10^{55}$~ergs corresponding to the
energy release from  $10^4$ to $10^5$ type II supernovae.
This fact suggests that they may have a common property and origin,
and can be modeled by a unified scheme of energy release in the galactic
center and its propagation into the halo and intergalactic space.

In this paper, we simulate  the extended soft X-ray halo
components in various galaxies based on the bipolar hypershell (BHS) model
developed for the Milky Way halo (Sofue 1977, 1984, 1994, 2000).
We further propose to use the simulation to
date and measure the explosive events at the nuclei,
such as their starburst history.
The simulation can also be used to obtain information about
the galactic-halo-intergalactic interface in disk galaxies.
The X-ray absorption morphology near the galactic
plane can be used to probe the interstellar gas distribution in the
galactic disk.

\section{Soft X-ray Hypershells in Nearby Spiral Galaxies}

\subsection{NGC 253}

ROSAT images of the nearby ($d=2.6$~kpc, Puche \& Carignan 1988)
starburst galaxy NGC~253 revealed
diffuse X-ray emission from the nuclear region, the disk and the halo
hemispheres (Pietsch et al. 2000). The emission measures of the three
components are nearly identical, and the integral diffuse emission
($L_{\rm x} = 4\times 10^{39}$~erg~s$^{-1}$) accounts
for $\sim 80\%$ of the total X-ray luminosity from NGC~253.
The diffuse bulge
emission is due to an extended, highly absorbed source with an extent of
250~kpc, and a hollow-cone shaped ``X-ray plume'' with an extent of
$\sim 700$~pc along the SE minor axis. The X-rays in the bulge region
are thought to trace the interaction between a galactic super-wind driven by
the starburst nucleus and dense interstellar medium in the disk.

In the disk, the highly absorbed
diffuse emission follows the spiral arm structure.
It most probably reflects a superposition of diffuse X-ray emission from hot
interstellar and discrete sources, such as emission from binaries, SN-remnants or
{\sc Hii} regions.
Two major components of the diffuse halo emission can be separated: ``coronal
emission'' arising from hot interstellar medium close to the disk (scale
height $\sim 1$~kpc) and emission from the ``outer halo''. NGC~253 is not
completely oriented edge-on, and the NW side of the galaxy is closer to us.
Due to this viewing geometry and absorption of the soft X-rays by the disk of
NGC~253, the coronal emission can only be detected from the near (SE) side of
the galaxy, while emission from the outer halo component is partly shielded
by the {\sc Hi} disk of NGC~253 to the NW. Following Pietsch et al. (2000),
the X-ray corona might be fueled by hot gas from galactic fountains
within the disk, while the outer halo emission is attributed to a superwind
emanating from the starburst nucleus. The latter assumption can also explain
the ``hour-glass-'' or ``dumbbell-shaped'' morphology of the outer X-ray halo.
Pietsch \& Vogler (2000) estimate  temperature of the outer halo component
at 0.2 keV
(2.3 million K), corresponding to an equivalent shock-heating velocity of
200 \kms.

\subsection{NGC 1808}

NGC 1808 is a galaxy with
large-scale cylindrical dusty jets emanating from the
nucleus at high velocities, and is supposed to be a mildly tilted M82 type
galaxy (V\'eron-Cetty and V\'eron 1985; Phillips 1993).
Soft X-ray imaging observations with the ROSAT PSPC and HRI revealed an active
X-ray nucleus and extended emission in the disk region
(Junkes et al. 1995; Dahlem et al. 1994), as well as a very extended X-ray
emission in the halo (Iwaki et al. 1996).
The total soft X-ray luminosity is of the order of $3\times 10^{40}$ ergs s$^{-1}$.
Iwaki et al. (1996) obtained wide area ASCA images in the 0.5--2 keV band,
and found that the extended component is highly elongated in the direction
of the dusty jets on both sides of the galactic plane.
Furthermore, the X-ray halo appears to be bifurcated into two ridges,
suggesting that it comprises double-horn feature or a partial shell structure.

\subsection{NGC~3079}

The nearly edge-on galaxy NGC~3079  is known to host an
active nucleus  classified as a LINER or Seyfert {\sc ii} nucleus, and
giant radio lobes and H$\alpha$ superbubbles emanate along the minor axis
(Irwin \& Seaquist 1991; Cecil et al. 1995).
ROSAT observations (Pietsch et al. 1998) detected an
integral 0.1--2.4~keV luminosity of $L_{\rm x} = 3\times 10^{40}$~erg~s$^{-1}$,
which can be resolved into three components. One component consists of
extended emission from the
central region ($L_{\rm x} = 1\times 10^{40}$~erg~s$^{-1}$) originating from
the region of the central superbubble. While most of this emission is thought
to represent hot gas in the superbubble, one cannot exclude additional
contributions of $\le 30\%$ from the active nucleus.
The second main
emission component arises from the disk of the galaxy
($L_{\rm x} = 7\times 10^{39}$~erg~s$^{-1}$) and can be partly resolved into
point-like sources. Each of the three detected point sources has
$L_{\rm x} \sim 6\times 10^{38}$~erg~s$^{-1}$.
Finally, emission from the halo ($L_{\rm x} = 6\times 10^{39}$~erg~s$^{-1}$)
is detected. As in the case of NGC~253, the hot gas does not fill the
halo hemispheres uniformly, but an hourglass-shaped structure of the halo emission is
reported. A mean temperature for the halo gas was estimated to be $3.5\times
10^6$~K.
Interestingly, the lowest contours of the extended emission show a saddle shape,
with a neck at the galactic plane, suggestive of bipolar double-horns.

\subsection{NGC~4258}

NGC 4258's anomalous arms could also be due to out-of-plane
partial shells (van Albada and van der Hulst 1982; Vogler and Pietsch 1999b).
Diffuse X-ray emission of the
galaxy was detected with  $L_{\rm x} \sim 2\times 10^{40}$~erg~s$^{-1}$ and
fills $\sim 40\%$ of the $D_{25}$ diameter of NGC~4258. At least one half of
the diffuse emission is caused by the anomalous spiral arms of this galaxy,
and the X-ray emission morphology closely follows the pattern detected in the
radio wavelengths.

In the literature, those arms are discussed to be
(a) jets confined to the disk of the galaxy and
expelled by the active nucleus (e.g., Martin et al. 1989, Hummel et al. 1989,
Plante et al. 1991, Cecil et al. 1995),
(b) outflows (jets or plumes) from the
active nucleus leaving the plane of the galaxy and possibly `raining down'
onto the disk again (van der Kruit et al. 1972), (c) a hypershell of hot
interstellar medium as present in the nuclear region of NGC~253 and
expanding from the disk into the halo (Sofue 1984, 2000), or (d)
bar shocks heating the interstellar medium due to the high differential
velocities between the bar orbit and interstellar medium in the bulge and
inner disk of NGC~4258 (e.g., Cox \& Downes 1996).

Remaining diffuse emission not correlated with the anomalous arms
might originate from the disk and the halo of the galaxy.
For a possible halo component, the temperature was estimated to be
$2\times 10^6$~K. This temperature is lower by a factor of 2 than the
temperature of the hot gas along the anomalous spiral arms
($T\ge4\times 10^6$~K).

In addition to the diffuse emission, 14 point sources
(integral $L_{\rm x} \sim 2\times 10^{39}$~erg~s$^{-1}$) were found in the
galaxy. For possible contributions of the active nucleus to the X-ray emission,
an upper limit of $L_{\rm x} \sim 1\times 10^{38}$~erg~s$^{-1}$ was
found in the ROSAT 0.1--2.4~keV band.

\subsection{M~82}

M82 is an edge-on starburst galaxy, which ejects a galactic-scale  flow through
bipolar cylindrical jets (Nakai et al. 1987) to which
larger-scale X-ray extended features are associated (Lehnert et al. 1999).
Strickland et al. (1997) report the detection of a galactic wind.
The X-ray emission was visible out $\ge 6$~kpc
above and below the plane of the galaxy, and the luminosity of the diffuse
X-ray emission outside the nuclear region is
$\sim 2\times 10^{40}$~erg~s$^{-1}$. From X-ray spectra, the temperature
of the gas is found to vary from
0.6~keV near the nucleus to 0.4~keV for distances
further away from the plane. The authors compared the X-ray results with
several theoretical models, namely Tomisaka \& Ikeuchi (1988), Tomisaka \&
Bregman (1993) and Suchkov et al. (1994). From their comparison, they favor
the model of Suchkov et al. (1994). This model assumes a two component cold
rotating dense disk, a non-rotating hot tenuous halo, and a starburst history
incorporating the milder mass and energy input from stellar winds into the halo
before more energetic supernova explosions dominate the scenario. Shocked halo
gas then provides the major emission component in the ROSAT band.

A reanalysis of the ROSAT
data (Lehnert et al. 1999) and new H$\alpha$ images showed the existence of a
4~kpc $\times$ 1~kpc region of spatially coincident X-ray and H$\alpha$
emission about 10~kpc north of the nucleus. The luminosity in both
wavebands accounts for roughly 1$\%$ of the total M~82 luminosities in both
bands. The authors attribute the H$\alpha$ emission to ionizing radiation
from the starburst that propagates into the halo, and the
X-ray emission to a shock-heated massive ionized cloud in the halo, which is
encountered by the superwind. With this scenario, they follow theoretical
predictions of  Suchkov et al. (1996).

\subsection{The Milky Way}

Using ROSAT All Sky Survey (RASS) data, presented in several energy
bands from 0.1 to 2.4 keV in Snowden et al. (1997), we showed that the
North Polar Spur (NPS) and its western and southern counter-spurs
trace a giant dumbbell-shape wit a neck at the galactic plane (Sofue 2000).
These features are interpreted as effects of a shock front originating from
a starburst 15 million years ago with a total released energy of the
order of $\sim 10^{56}$ ergs or $10^5$ type II supernovae.
We simulated all-sky distributions of radio continuum and soft X-ray
intensities based on the bipolar hypershell galactic center starburst model.
The simulations nicely reproduce the radio NPS and related spurs, as well as
radio spurs in the tangential directions of spiral arms.
Simulated X-ray maps in 0.25, 0.75 and 1.5 keV bands reproduce the
ROSAT X-ray NPS, its western and southern counter-spurs, and the
absorption layer along the galactic plane.
We further proposed to use the ROSAT all-sky maps to probe
the physics of gas in the
halo-intergalactic interface of the Milky Way, and to directly date and measure
the energy of a recent Galactic Center starburst.
The present paper is a series of the BHS modeling of extended halo X-rays in
galaxies using the same numerical code.

\section{Bipolar Hypershell Starburst Model}

We adopt the same numerical methods to trace the shock envelope and
X-ray intensity distributions as we used to model the radio spurs and
BHS in the Milky Way, which is described in detail in Sofue (2000).
We briefly describe the methods below.

\subsection{Adiabatic-Shock Envelope Method}

The propagation of a shock wave through the galactic halo induced by a point
energy injection at the center can be calculated by applying the
shock envelope tracing method of Sakashita (1971)  and  M{\"o}llenhoff  (1976),
who extended the Laumbach and Probstein (1969) method
for tracing the evolution of a shock front to a case of axi-symmetric
distributions of ambient gas.
The flow field is assumed to be locally radial, and the gas is adiabatic,
and, therefore, the heat transfer by radiation and counter-pressure are
neglected.
The density contrast between the shock front and ambient gas is given
by $(\gamma+1)/(\gamma-1)=4$ for $\gamma=5/3$, where $\gamma$ is the adiabatic
exponent of the gas.

The equation of motion  of the shock wave is given as follows
(M{\"o}llenhoff 1976):
$$
E_0 = \int^R_0 {P \over{\gamma-1}} 4 \pi r^2 dr
+ \int^R_0 {1 \over 2} {\left( {\partial r \over \partial t} \right) ^2}
\rho_0 4 \pi r_0^2 dr_0					\eqno(1).
$$
Here, $E_0$ is the total energy of the explosive event,
$P$ is the internal pressure, $\gamma$ is the adiabatic exponent,
$\rho_0$ is the unperturbed ambient gas density, and $r$ is the radius
from the explosion center, with suffix 0 denoting the quantities of the
unperturbed ambient gas, and $R$ is the radius of the shock front.
This equation then leads to an equation of the shock radius, as described
in Sofue (2000).

The unperturbed density distribution of gas in a galaxy is
assumed to  comprise a stratified disk, a halo with an exponentially
decreasing density, and intergalactic gas with a uniform density.
The differential distribution is approximated by the following expression.
$$
\rho_0= \rho_1 {\rm exp}(-(z/z_1)^\eta))
+ \rho_2 {\rm exp}(-z/z_2)
+ \rho_3 						\eqno(2).
$$
Here, suffices 1, 2 and 3 denote quantities for the disk, halo
and intergalactic  gas, respectively,
$\rho$ is the density, $z$ is the height from the galactic plane,
$z_i$ is the scale thickness of the disk and halo.
The power index $\eta$ represents the tightness of the disk component toward
the galactic plane, and is taken either 1 (exponential) or 2 (gaussian).
The third component, representing the intergalactic gas density, is taken
to be constant at $\rho_3=10^{-5} {\rm H~cm}^{-3}$, while the other parameters
are set in various possible combinations as listed in Table 1.
We also calculate a case in which the halo density decreases with a power law.
We further calculate another case with uniform halo density which decreases suddenly
at a certain height facing the intergalactic gas, which represents a case
for halo gas confined by the intergalactic gas pressure.


\begin{table*}
\caption{Parameters for Shock Envelope Solutions}
\begin{tabular}{cccccccc}
\tableline \tableline
 Model& Energy & $\rho_1$ & $h_1$ & Disk $z$ type& $\rho_2$ & $h_2$ & $\rho_3$ \\
    &  ergs &  H cm$^{-3}$ &  kpc &       & H cm$^{-3}$ & kpc & H cm$^{-3}$ \\
\tableline
 S1   & $0.4\times 10^{55}$
	 & 1   &  0.1 &exponential & 0.01  & 1 & $10^{-5}$   \\
 S2   & "& 1   &  0.2 &exponential & 0.01  & 1 & "  \\
 S3   & "& 1   &  0.1 &gaussian & 0.01  & 1 & "  \\
 S4   & "& 1   &  0.2 &gaussian & 0.01  & 1 & "  \\
 S5   & "& 1   &  0.1 &power law & 0.01  & 1 & "  \\
 S6   &  $1.6\times 10^{55}$
	 & 0.1   &  2 &exp[$-(z/h_1)^4$] & 0.0  & -- & "  \\
\tableline
\label{table1}
\end{tabular}

The shock envelope (radius and velocity) obeys a scaling law: For equal
ratios $E/\rho_i$, one obtains the same envelope, independent of the
individual $E$, $\rho_i$ values

\end{table*}
Solutions of the above equations are presented  in Fig. 1
for various parameter sets given in Table 1.
The total energy input is taken to be $E_0=0.75\times 10^{55}$ ergs for S1 to S5,
and $3\times 10^{55}$ ergs for S6, and the shock
envelopes are drawn every 10 million years.
The shock velocity can be estimated by dividing the radial distance
between succeeding envelopes by $10^7$ years.

Models S1 and S2 : Exponential disk with scale height 0.1 and 0.2 kpc,
respectively, exponential
halo with scale height 1 kpc, and intergalactic gas of uniform density.

Models S3 and S4 : Gaussian disk with half thickness 0.1 and 0.2 kpc,
respectively, exponential disk of scale height 1 kpc, and intergalactic
gas of uniform density.

Model S5: Power-law density disk, power-law density halo, and intergalactic
gas of uniform density.

Model S6: Disk of 2 kpc thickness with sudden decrease of density,
facing the intergalactic low-density uniform gas.
This model represents a case where the halo-to-disk gas is not
gravitationally stratified (e.g., as exponentially),
but is confined by intergalactic gas pressure, so that the internal
pressure and density in the halo are more uniform.

--- Fig. 1 ---

\subsection{X-ray Emission and Absorption}

The X-ray intensity distribution for a model galaxy having a
hypershell is calculated as below.
The centers of the shells are assumed to be at $z=\pm6$ kpc on the rotation
axis, and the radii 6 and 9 kpc in the radial and vertical directions,
respectively.
The volume emissivity is calculated from the
density contrast of the shocked gas,   which  is assumed to be
proportional to $\rho/\rho_0=(\gamma+1)/(\gamma-1)$,
where $\rho$ and $\rho_0$ are gas densities in the shocked shell and
unperturbed halo gas, respectively.
In the present simulation, the profile of emissivity perpendicular to the shell
surface is simply represented by an exponentially decreasing function
behind the shock front toward the center with a scale thickness of 500 pc.
The emissivity also decreases with the height from the galactic
plane with a scale height of 3 kpc, corresponding to the exponentially
decreasing gas density in the halo.

The X-ray emission from a hypershell is assumed to be
thermal free-free radiation, whose emissivity is given by
$$
\epsilon = n_{\rm e}^2 \Lambda
\propto \rho^2 T^{1/2} 					\eqno(3).
$$
Here, $\Lambda$ is the cooling function of the gas.
Since the temperature is as high as $T\sim 10^7$ K in our simulation,
the radiation is almost totally free-free, and the contribution by
recombination lines from hydrogen, helium, and metals, which are
significant at $T\sim 10^5$ K, can be neglected.
A typical emission measure along the hypershell ridge is
$n_{\rm e}^2 L \sim 10^{-2}$ cm$^{-6}$pc for a tangential pass in the shell
of about 2 kpc.
For simplicity, we assume that the temperature of X-ray emitting gas
is  constant in the shocked shell.
In addition to the hypershells,
we assume a galactic disk X-ray component of scale height of 500 pc, and
a bulge component of scale radius 1 kpc.
X-ray emissions from the  ambient (unperturbed) halo gas
is neglected, because the density would not be sufficiently high
to contribute a significant 0.1-2.4 keV X-ray emission measure.

The X-ray intensity, $I$, is calculated by
$$
dI= \epsilon_{\rm X} ds - \kappa I ds			\eqno(4),
$$
where the first term in the right-hand side is the emission measure with
$\epsilon_{\rm X}$ being the X-ray emissivity,
and the second term represents the absorption rate with
$\kappa$ being absorption coefficient,
and $s$ is the distance along the line-of-sight.

The interstellar extinction of soft X-rays occurs due to
the photoelectric absorption by metals, and has been
calculated for the solar metal abundance
by Morrison and McCammon (1983).
The absorption coefficient $\kappa$ is, then, represented by
$$
\kappa = (n_{\rm H}/N_{\rm H}^0) (E^{\rm X}/E^{\rm X}_0)^{-2.5} \eqno(5),
$$
where $n_{\rm H}$ is the number density of hydrogen atoms,
and $N_{\rm H}^0$ is the $e$-folding column density of interstellar
neutral hydrogen at the photon energy $E^{\rm X}_0$
(Ryter 1996; Sofue 2000).
At $E^{\rm X}_0=0.75$ keV, the $e$-folding column density for
$E^{\rm X}_0=0.75$ keV
is given as $N_{\rm H}^0=3 \times 10^{21}$ H cm$^{-2}$ (Snowden et al. 1997).
Since the ROSAT energy bands are rather broad, covering
higher and lower energies around the representative energies,
we adopt two representative cross sections
at 0.75 and 1.5 keV bands, which are 1 and 0.1 times the value at
0.75 keV, respectively.

The absorbing disk gas  is assumed to be condensed in logarithmic spiral arms
of pitch angle 6$^\circ$ and the width is taken to be
1/5 of the arm separation, as in the same way assumed for the Milky Way.
The peak hydrogen density in an arm at a distance of
8 kpc from the center  is taken to be 5 H cm$^{-3}$.
The hydrogen density distribution is expressed by
$$
n_{\rm H}= \alpha n_0 {\rm exp}(-r/r_H -z/z_H)
 {\rm cos}^k (\theta - \eta {\rm log}~r/r_0)			\eqno(6).
$$
Here, $\theta$ and $\eta$ are the azimuthal angle and the pitch angle
of the arms.

\section{Results of Simulations}

\subsection{NGC 253}

The expansion velocity of the outer halo is estimated to be 200 \kms,
which is required for shock heating of the gas up to 0.2 keV ($2.3 \times 10^6$ K)
(Pietsch et al. 2000).
Taking this expansion velocity for the BHS at 5 kpc height from the galactic plane
in the ROSAT 0.75 keV contour map, we find that the observation is well
fitted by a shock front at $\sim 3\times10^7$ yrs for an explosion energy
of $\sim 4 \times 10^{54}$ ergs as in Table 1.

Among the models, however, a better fit is obtained by Models S2 and S6
than by the other models, which indicates
that the decrease of halo gas density in the direction
perpendicular to the disk is not as steep as in the other models.
In Fig. 2 we show the shock front for Models S2 and S6
superposed on the 0.75 keV ROSAT map of NGC 253 in contours
overlaid on a B-band optical photograph from DSS.
The ROSAT result shows that the hypershells tend to break and become
round at high altitudes, when the shock front reaches a few kpc
from the galactic plane, mimicking a dumbbell shape.
Hence, Model S6 appears to be preferable to S2, for which the BHS is
more like an ``hourglass".
Model S6 represents a case in which the gas density is uniform
within the disk and halo up to a certain height (e.g. 2 kpc),
where the halo faces a uniform intergalactic gas of lower density.
Such density distribution will be possible if the halo gas is confined by
the intergalactic gas pressure, but not stratified by a
hydrostatic equilibrium under the disk gravity.

---- Fig. 2 ----

Although Model S6 appears better, overall reproduction of the
observations is obtained by both models.
For simplicity, we hereafter use the Model S2 in which the shape of the
shell can be well approximated by a part of an ellipsoidal sphere.
This allows us a precise prediction for the X-ray emissivity and density
distributions.
Fig. 3 shows the calculated intensity distributions for NGC 253
for two different opacity ($\kappa$) values, which correspond to
(a) a case of 0.75 keV band, and (b) 1.5 keV band.
We also show the ROSAT X-ray images in the right panels: right-top for
0.75 keV in gray scale, and the right-middle and right-bottom are for
0.75 and 1.5 keV, respectively, overlaid on an optical photograph.
As seen in the figures, the BHS simulation reproduces
the following observed characteristics:
(1) double-horn morphology of the diffuse emission features in
the 0.75 keV band emerging from the disk,
(2) strong absorption along the galactic plane, and
(3) energy dependence of the intensity distribution, namely, harder emission
near the disk.

--- Fig. 3 ---

The Combination of three NGC 253 images in independent energy bands centered at
0.25, 0.75 and 1.5 keV enabled the construction of a true color image
(Pietsch et al. 2000).
For this purpose, the different subbands were associated with the colors
red, green and blue, for the 0.25, 0.75 and 1.5 keV band, respectively.
Fig. 4 shows the thus obtained colored intensity distribution compared to
the observed spectral intensity distribution taken from Pietsch et al. (2000),
where red stands for soft and blue for hard.
The present BHS model can nicely reproduce the observed spectrum distribution,
such as the harder spectra in the disk and spiral arms due to stronger
absorption of softer X-rays.

--- Fig. 4. ---

\subsection{BHS at Various Inclination: Comparison with Other Galaxies}

We have thus shown that the extended X-ray
features in the halo of NGC 253 are well reproduced using the BHS
model, which was originally used to model the North Polar Spur NPS and
associated spurs in the Milky Way halo (Sofue 2000).
It is, therefore, interesting to see, whether the BHS model also applies to
extended X-ray features observed in other galaxies exhibiting starbursts
and outflows.
Here, we compare the observed morphology with the outcomes of our
theoretical description.
We calculated 0.75 keV intensity distributions
for various inclination angles at 0\deg (face-on), 30\deg, 45\deg,
60\deg, 80\deg, and 90\deg (edge-on).
Fig. 5 shows the calculated results, and they are compared in Fig. 6 with
observed examples for
M83 ($i=24^\circ$),
NGC 1808 ($i=58^\circ$),
NGC 4258 ($i=67^\circ$),
NGC 253 ($i=78^\circ$),
NGC 3079 (edge-on),
M82 (edge-on)
and the Milky Way (edge-on).

--- Fig. 5 ---

--- Fig. 6 ---

{\it M83}: When a galaxy with BHS is observed face-on,
the X-ray images exhibit a bright ring feature as in Fig. 5.
If the galaxy is slightly inclined such as
$i=30\Deg$, the BHS ridges in both sides of the galactic plane
are superposed to exhibit a box-shaped bright region.
Such a box-shaped X-ray distribution appears to be observed indeed in
the ROSAT PSPC image of M83, as shown in Fig. 6 (Strickland et al. 1997),
where the arrow points toward the direction of the minor axis as the near side.
Although it is difficult to distinguish the halo emission from disk
contributions, spiral arms and/or the bar, some emission along the
boxy edges could be due to overlapped ridges of BHS.

{\it NGC 1808}: The extended 0.5--2 keV features bifurcated in the halo of
NGC 1808 as observed with ASCA (Iwaki et al. 1996)
are also reproduced by our simulations,
although the lopsidedness cannot be reproduced well.
The arrow in Fig. 6 indicates the direction of the dusty jet along the
minor axis in the nearer side.

{\it NGC 4258}:
This  is a galaxy known for its anomalous radio arms, along which X-ray
extended emission is observed, as is shown in the 0.1--0.4 keV ROSAT map
(Vogler and Pietsch 1999b).
Although the resolution is still too crude,
some spur features are mimicked by the BHS model.
The origin of the anomalous arms is controversial, as described in
section 2, but we comment that the anomalous radio arms
(van Albada and van der Hulst 1982) are nicely fitted by two oppositely
extending ridges of the BHS.
In particular, their round shapes and curvature of the arms returning to
the disk can be explained if they are part of BHS.
However, the lobsidedness of the anomalous arms with respect to the
rotation axis remains a question in the BHS model, although some
lopsidedness is seen in the BHS of the Milky Way (Fig. 6 bottom, constructed
from the energy bands presented in Snowden et al. 1997).

{\it M82}: The ROSAT HRI image (Strickland et al. 1997)
shows a bipolar extended halo elongated in the direction perpendicular to
the disk, and the emission appears to be centrally filled, indicating that
the outflow is well collimated, as well simulated by the ``jet-like wind''
model by Suchkov et al. (1994, 1996).
We mention, however, that the northern end of the flow at a height of
about 2.5 kpc (3$'$)
from the disk appears to be more opened into the intergalactic space,
suggesting double-horn ridges or a partial shell.

{\it NGC 3079}: The radio-lobe edge-on galaxy NGC 3079
shows a box-shaped X-ray features, suggesting a double-horn morphology,
as observed in the ROSAT PSPC image (Pietsch et al. 1998),
and is well reproduced by an edge-on BHS model.

{\it Milky Way}: As mentioned in Sect. 2.6,
the RASS
reveals bipolar hypershells, which make up the North Polar Spur and its
counter-spurs. The observations are well fitted by the BHS model.
A detailed modeling for the case of the Milky Way is described in Sofue (2000).

\section{Discussion}

\subsection{Probing Starburst and Halo Gas Distribution}

In the present shock approximation, the expansion velocity $v$ at radius $r$
is related to the mean ambient density $\rho_0$ and the explosion energy $E_0$ as
$E_0 \sim \rho_0 r^3 0 v^2$.
Here, $r$ is measured from the observed shell radius, and
the expansion velocity is also measured from the
gaseous temperature inferred from X-ray spectrum.
For given radii and velocities (temperature), which are both
known parameters from observations, a scaling law can be established:
In the case of constant $E / \rho$ values, we obtain identical shock envelopes.

For a given temperature (velocity), the emission measure is estimated
from the intensity, and therefore, the accumulated total amount of gas
in the shell is known, which is related to the mean gas density $\rho_0$.
Hence, we can measure $\rho_0, ~ r$ and $v$ from observations,
and thus, we can finally estimate the total energy of the explosion.

The morphology of the BHS is sensitive to the density
distribution in the disk and halo.
The shell shape in the uppermost part manifests the halo-intergalactic
density structure.
If the intergalactic gas density is very low, the shell will have a
larger on-average
radius and a more conical form. In the case of higher density,
the shell becomes compact and round.
Among the various ambient density distributions in Table 1, Models S2 and S6
appear to better fit the observations than the other models.
Although both models allow a sufficient global reproduction of the
observed morphology of NGC 253, Model S6 appears to be preferable to
reproduce the dumbbell shape of the observed X-ray lobes.
This suggests that the disk-to-halo gas density is rather uniform until it
faces the intergalactic low-density region.
Such uniform density distribution may be possible if the gas is confined
by the gas pressure, not hydrostatic equilibrium in the disk gravity.
In order to clarify this and fit the observations in more detail, higher
resolution imaging in the soft X-rays is necessary.

\subsection{Origin of the Hypershells}

In our model, the energy injection at the galactic center is assumed
to be impulsive, and the shock is strong enough to create a well-defined
shell structure.
The required total energy given to the interstellar gas is of the order
of $\sim 10^{55}$ ergs.
The time scale of the explosive event responsible for the energy
release is assumed to be significantly shorter than
the expansion timescale of the hypershell,
$t\sim r/v \sim 10^7$ years for $r\sim$~several kpc and $v\sim300$ \kms.
So, the explosion (starburst) time scale must be shorter than a few million years.
Such impulsiveness and robustness of the energy release
can be explained if the nucleus of the galaxy has experienced a starburst
$\sim 10^7$ years ago, lasting for a few million years or shorter,
during which $\sim 10^{4...5}$ type II supernovae exploded. It is worth
mentioning that similar timescales are predicted from starburst modeling based
on infrared observations
(e.g., F\"orster-Schreiber et al. 2000).

Alternative models have also been discussed by several authors:
Suchkov et al. (1994, 1996) developed a mass-loading outflow model,
and simulated the X-ray jets in M82 using their hydrodynamical code.
In their model, the blowout itself is as hot or as highly energetic as one
would detect in the ROSAT band.
This model may provide a better fit in the case of M82, where we observe
a centrally-filled high energy
flow well collimated in the direction perpendicular to the disk.
A stellar-wind driven Galactic wind (Heckman et al. 1990) could
also produce a shell structure, as observed with ROSAT for
NGC 253 (Pietsch et al. 2000).
If the duration of the starburst is short enough compared to the expansion
time scale of the shell, namely, if the starburst
is ``impulsive'' compared to the shell's life time, the wind model and the
our shocked BHS model give essentially the same result.
On the other hand, if the wind is steady and longer-lived than the
shell's expansion time, e.g. a few tens of million years,
the flow would become an open-cone shape, which does not apply to
round shells observed in the case of NGC 253.

\section{Summary and conclusions}

We presented a theoretical model of the million degrees phase of the
interstellar medium in the bulge, disk and halo of spiral galaxies. In this
model, we assume a sudden energy release in the nuclear region
of the galaxy, as provided, e.g., by a nuclear starburst.
The energy release is supposed to form a bipolar hyper-shell
(BHS) of ionized ISM, which is heated to a million degrees and is thus visible in
the soft X-ray band. Our simulations are
based on the assumption of an adiabatic shock envelope, and our projections
of the three-dimensional data cubes for different galaxy inclinations take into
account the absorption of soft X-rays by intervening cold gas in the disk of
the galaxy. The model predicts different spectral energy
distributions of the soft X-ray emission for different distances from the
galactic plane.

We applied our model to a
mini-sample of nearby spiral galaxies, which are all classified as starburst
galaxies, and the model-predictions are nicely supported by the observations.
 In the case of the very nearby galaxy NGC~253 ($d=2.6$~Mpc),
the X-ray images
allow a high spatial resolution. Our model reproduces the ROSAT observations,
namely the hourglass-shaped morphology, the absorption of disk and halo
emission from the reverse (NW) side of the nearly edge-on galaxy due to
cold gas in the disk, as well as the spectral distribution of the emission
components.

We extended our model to other spiral galaxies
seen under different inclinations, edge-on (NGC~3079, M~82),
under high inclination (NGC~4258, NGC~1808) or face-on (M~83). As for NGC~253,
our simulations match the observed X-ray morphology and spectral energy
distribution. The X-ray corona of the Milky Way, observed in the ROSAT All Sky
Survey, can also be matched with our model. This topic is briefly
discussed in this publication and is subject of another, more detailed
publication (Sofue 2000).

From the comparison of our simulations with the observations, we are able
to deduce expansion velocities of the shells between 200 \kms and 300 \kms,
which are necessary to heat the gas to $(2-4)\times 10^6$~K. The age of
the shells has to be of the order of some $ 10^7$ years, and from this
results we can derive an upper limit for the duration and intensity of the
initial starburst which provided the energy release to form the BHS
morphology. In general, this energy release is estimated to be of the order
of $10^{55}$~ergs for all galaxies, and typical timescales of the burst seem
to be $\sim$ several $10^6$ years. Such timescales are also predicted
from infrared observations in the case of M~82, NGC~253 and NGC~1808
(e.g., F\"orster-Schreiber et al. 2000).

The new generation X-ray observatories, {\it Chandra} and XMM, both now
in orbit, have larger energy bands than ROSAT had
(0.2--10~keV versus 0.1--2.4~keV) and offer superior spectral resolution.
They allow us
to study spiral galaxies at sub-arcsec resolution ({\it Chandra})
and to probe the ISM with an extremely high photon collecting power
at a resolution of $\sim 10''$ (XMM). The
combination of {\it Chandra} and XMM results will provide spatially highly
resolved images of the bright ISM features, as well as detailed X-ray
spectra of the ISM in general.
The BHS model, in which the shell morphology is sensitive to the
ambient density distribution, will then allow us to
probe the distributions of the million degree interstellar medium
in the disk, halo and intergalactic space as well as the disk halo interface
at high spatial and spectral resolution.

\vskip 2mm
We are grateful to W. Pietsch for helpful comments and
providing the NGC~3079 and NGC~253 data. We would like to thank Sue Madden
for carefully proofreading the manuscript.

\vskip 2mm

\def\bibitem{\r}
\noindent{\bf References}

\bibitem{} Cecil G., Morse J.A., Veilleux S., 1995, ApJ 452, 316

\bibitem{} Cox P., Downes D., 1996, ApJ 473, 219

\bibitem{} Dahlem M., Hartner G. D., Junkes N., 1994, ApJ 432, 598

\bibitem{} F\"orster-Schreiber N., et al., 2000, A\&A, in prep.

\bibitem{}  Haslam C.G.T.,  Salter C.J.,  Stoffel H.,  Wilson W.E.,  1982,  A\&A
Suppl. 47,  1

\r\bibitem{} Heckman T. M., Armus L., Miley G. K., 1990, ApJS 74,  833

\bibitem{} Hummel E., Krause M., Lesch H., 1989, A\&A 211, 266

\bibitem{} Immler, S., Vogler, A., Ehle, M., Pietsch, W. 1999,  A\&A 352, 415

\bibitem{} Irwin J.A., Seaquist E.R., 1991, ApJ 371, 110

\bibitem{} Iwaki H., Ueno S., Koyama K., Tsuru T., Iwasawa K., 1996, PASJ 48, 409

\bibitem{} Junkes N., Zinnecker H., Hensker, G. Dahlem M., and  Pietsch W.,
1995, A\&A 294, 8

\bibitem{} Laumbach D. D., Probstein R. F., 1969, J. Fluid Mecha. 35, 53

\bibitem{} Lehnert M. D., Heckman T. M., Weaver K. A.,
1999, ApJ 523, 575 

\bibitem{} Martin P., Roy J.-R., Noreau L., Lo K.Y., 1989, ApJ 345, 707

\bibitem{} M{\"o}llenhoff C., 1976, A\&A 50, 105

\bibitem{} Morrison R., McCammon D., 1983, ApJ 270, 119

\bibitem{} Nakai N., Hayashi M., Handa T., Sofue Y., Hasegawa T., Sasaki M.,
1987, PASJ 39, 685

\bibitem{} Phillips A. C. ,1993, AJ 105, 486.

\bibitem{} Pietsch W., Trinchieri G.,  Vogler A., 1998, A\&A 340, 351.

\bibitem{} Pietsch W., Vogler A., Klein U., Zinnecker H., 2000,
 A\&A 360, 24

\bibitem{} Plante R.L., Lo K.Y., Roy J.-R., Martin P., Noreau L., 1991, ApJ 381, 110

\bibitem{} Puche D., Carignan C., 1988, AJ 95, 1025

\bibitem{} Ryter Ch. E., 1996, ApSpSc 236, 285

\bibitem{} Sakashita S., 1971, ApSpSc 14, 431

\bibitem{}  Snowden S. L., Egger R., Freyberg M. J., McCammon D.,Plucinsky P. P.,
  Sanders W. T., Schmitt J. H. M. M., Tr\"umper J., Voges W. H.,
1997, ApJ 485, 125

\bibitem{}  Sofue Y., 1977, A\&A 60, 327 

\bibitem{}  Sofue Y., 1984,  PASJ  36,  539

\bibitem{} Sofue Y.,  1994, ApJ.L., 431, L91

\bibitem{} Sofue Y., 2000 ApJ 540, 224.

\bibitem{} Strickland D. K., Ponman T. J., Stevens I. R., 1997, A\&A 320, 378.

\bibitem{} Suchkov A.A., Dinshaw S.B., Heckman T.M. Leitherer C., 1994, ApJ 430, 511.

\bibitem{} Suchkov A.A., Berman V.G., Heckman T.M., Balsara D.S., 1996, ApJ 463, 528

\bibitem{} Tomisaka K., Ikeuchi S., 1988, ApJ 330, 695

\bibitem{} Tomisaka K., Bregman J. N., 1993, PASJ 45, 513

\bibitem{} van Albada G.D., van der Hulst J.M., 1982, A\&A 115, 263

\bibitem{} van der Kruit P.C., 1974, ApJ, 192, 1

\bibitem{} Veron-Cetty M. P., Veron P., 1985, A\&A 145, 425

\bibitem{} Vogler A., and Pietsch W., 1999a, A\&A 342, 101 

\bibitem{} Vogler A., Pietsch W., 1999b, A\&A 352, 64

  \vskip 10mm

\noindent{\bf Figure Captions}\vskip1cm

\r Fig. 1: Calculated shock front in the galactic halo at every
$2.5\times10^7$ yrs after an explosion and/or a starburst at the
nucleus with a total energy of $0.4 \times 10^{55}$ erg
for various parameters as models S1 to S6 in Table 1

\r Fig. 2: Calculated shock fronts for Model S2 and S6 overlaid on the
ROSAT 0.75 keV contours and optical photograph of NGC 253
from Pietsch et al. (2000)

\r Fig. 3: Simulated intensity distributions for NGC 253 at
0.75 and 1.5 keV (left panel) compared with ROSAT images at 0.75 and 1.5 keV
taken from Pietsch et al. (2000) (right panel)

\r Fig. 4: Addition of 0.25 (R), 0.75 (G) and 1.5 (B) keV intensities,
compared with that of NGC 253 (Pietsch et al. 2000)

\r Fig. 5: Simulated 0.75 keV views of a galaxy at various inclination
angles

\r Fig. 6: Comparison of the model with observed X-ray images
of several galaxies with various inclinations:
From top to bottom, M83, NGC 1808, NGC 253, NGC 4258, M82, NGC 3079,
and the Milky Way.
The arrows in M83 and NGC 1808 point the directions of the
minor axis in the near side.
See the text for references

\end{document}